# Structural and magnetic properties of Cr-diluted CoFeB

Yishen Cui[1,a)], Manli Ding[1,a)], S. Joseph Poon[1], T. Paul Adl[2], S. Keshavarz[3], Tim Mewes[3], Stuart A. Wolf[1,4], Jiwei Lu[4]

[1]Department of Physics, University of Virginia, Charlottesville, Virginia 22904, USA

[2]Micron Technology Inc., Manassas, Virginia 20110, USA

[3]Department of Physics and Astronomy, MINT Center, University of Alabama, Tuscaloosa, Alabama 35487, USA

[4]Department of Materials Science and Engineering, University of Virginia, Charlottesville, Virginia 22904, USA

[a)] Yishen Cui and Manli Ding contributed equally to this work.

**Abstract:**

The crystallization process and the magnetization of Cr diluted CoFeB was investigated in both ribbon samples and thin film samples with Cr content up to 30 at. %. A primary crystallization of *bcc* phase from an amorphous precursor in ribbon samples was observed when the annealing temperature rose to between 421 ˚C and 456 ˚C, followed by boron segregation at temperatures between 518 ˚C and 573 ˚C. The two onset crystallization temperatures showed strong dependences on both Cr and B concentrations. The impact of Cr concentration on the magnetic properties including a reduced saturation magnetization and an enhanced coercive field was also observed. The magnetizations of both ribbon samples and thin film samples were well fitted using the generalized Slater-Pauling curve with modified moments for B (-0.94 $\mu_B$) and Cr (-3.6 $\mu_B$). Possible origins of the enhanced coercive field were also discussed. We also achieved a damping parameter in CoFeCrB thin films at the same level as $Co_{40}Fe_{40}B_{20}$, much lower than the value reported for CoFeCrB films previously. The results suggest a possible advantage of CoFeCrB in reducing the critical switching current density in Spin Transfer Torque Random Access Memory (STT-RAM).

# Introduction

Spin Transfer Torque Random Access Memory (STT-RAM) has become increasingly important in the memory industry due to its exceptional scalability[1,2]. However, the critical switching current density $J_{c0}$ for free layer magnetization switching in magnetic tunnel junctions (MTJ) remains large, preventing the STT-RAM bit size from scaling down[2]. $J_{c0}$ can be reduced by lowering the magnetization of the MTJ free layer (FL) electrode. It is also critically related to the Gilbert damping parameter of the FL material[2,3]. $Co_{40}Fe_{40}B_{20}$ has been dominantly used as FL electrodes in MgO-based MTJs. It has a magnetization of ~1200 emu/cc along with a benchmark damping parameter at the level of $0.004^4$. Besides, the amorphous feature of as-deposited $Co_{40}Fe_{40}B_{20}$ films facilitates the formation of a pure *bcc*(001) oriented CoFeB/MgO interface after annealing[5], thus leading to a large coherent tunneling magnetoresistance (TMR)[6,7].

Cr has been considered as a promising substitute for Co/Fe in CoFeB to reduce the magnetization and thus the critical current density[8,9]. With a composition of $Co_{40}Fe_{32.7}Cr_{7.3}B_{20}$ $J_{c0}$ was reduced by a factor of four to $4.9 \times 10^5$ A/cm$^2$ compared to MTJs with $Co_{40}Fe_{40}B_{20}$[9]. Additionally, a *bcc* crystallization is expected in CoFeCrB due to the high insensitivity to the constituent miscibility of CoFeB alloys, such as in CoFeNbB[10] and other diluted-CoFeB[11]. However, there is still a lack of detailed information on the crystallization process and the magnetic properties of CoFeCrB, particularly for high Cr content. In this paper, both the structural and magnetic properties were studied as functions of Cr contents and annealing temperatures. In addition, ferromagnetic resonance measurements were performed on CoFeCrB films, showing a much lower damping parameter compared to the values reported in the previous literature[9].

**Experiment**

To prepare ribbon samples, ingots with compositions of $Co_{40}Fe_xCr_yB_z$ ($x + y + z = 60$ and y varied in the range of 0~30) were prepared by arc melting the mixture of high purity Fe, Co, Cr, and B elements in an argon atmosphere. Each ingot was re-melted four times to ensure the chemical homogeneity. Amorphous ribbons, typically about 1.25 mm wide and 25 μm thick, were produced from the ingots melted spun on a copper wheel rotating at ~30 m/s. Onset crystallization temperatures were determined by differential scanning calorimeter (DSC) which has a constant heating rate of 5 ˚C /min in a temperature range from 50 ˚C to 600 ˚C. As-spun ribbons were also isochronally annealed in evacuated quartz tubes for 1 hour at various temperatures up to 600 ˚C.

CoFeCrB films were deposited at room temperature using an RF magnetron sputtering system. Thermally oxidized silicon wafers were used as substrates. Two targets with different compositions ($Co_{43}Fe_{29}Cr_{10}B_{18}$ and $Co_{40}Fe_{18}Cr_{30}B_{12}$) were co-sputtered with different power ratios. By varying the powers applied to the two targets, the Cr content can be systematically changed from 10% to 30 %, along with a reducing Fe content. The structure of all the thin film samples was Ta (7nm)/CoFeCrB (30nm)/Ta (7nm). A Ta film was used as a seeding layer to improve the surface smoothness and as a capping layer to protect CoFeCrB films from oxidation. The films were annealed at temperatures from 350 ˚C to 450 ˚C in vacuum for 1 hour.

Chemical compositions of both ribbon and thin film samples were determined using inductively coupled plasma-mass spectroscopy. The microstructures of ribbons and films were characterized by a high resolution X-ray diffractometer (XRD) with Cu Kα radiation (Smart-lab, Rigaku Inc.), and transmission electron microscopy (TEM). The film magnetic behaviors were characterized using the VSM option in Quantum Design PPMS-6000. Ferromagnetic resonance (FMR) spectra were acquired using a broadband coplanar waveguide field-swept setup with frequencies up to 50 GHz. The frequency dependence on the linewidth was used to estimate an effective damping parameter[12].

**Results and discussions**

**A. Structure properties**

Fig. 1 shows the amorphous structure of as-spun $Co_{40}Fe_{32}Cr_{10}B_{18}$ ribbons observed using both XRD and TEM. The broad *2θ* scan indicates an amorphous feature near 45° as seen in Fig. 1(a). In Fig. 1(b), the bright-field TEM image shows an amorphous matrix without any presence of crystallites. The inset diffraction pattern shows a halo ring that is typical for an amorphous phase. The amorphous phase without long-range crystalline order is due to the high B content. The amorphous precursor in CoFeCrB is expected to facilitate the formation of a coherent *bcc*(001)-oriented interface with MgO in MTJs[5].

Fig. 2(a) shows the *2θ* scans of the ribbon samples with the same composition ($Co_{40}Fe_{32}Cr_{10}B_{18}$) annealed at 417 °C and 508 °C respectively. A *bcc* phase with diffraction peaks at 44.95°, 65.56° and 82.94° were observed in the sample annealed at 417 °C, corresponding to peaks (110), (200), (211). The lattice constant is 2.85 Å close to the reported values for CoFeB[13] and FeCr[14]. As the annealing temperature was increased to 508 °C, additional peaks were observed and indexed as the boride-type phase $(Fe,Co)_3B$, which was similarly observed in CoFeB and CoFeNbB[10,13].

The DSC measurements for the same composition showed a good agreement with the XRD scans (Fig. 2(b)). Two exothermic peaks were observed, indicating a two-stage crystallization process as reported in CoFeB/diluted-CoFeB systems[10,11,13]. The two onset crystallization temperatures ($T_x$) could be determined accordingly. A primary *bcc* crystallization occurred at ~421 °C, followed by the boron segregation phases at a higher temperature of ~ 518 °C. Based on the DSC measurements, the onset crystallization temperatures as a function of the ribbon composition are summarized in Table I. It is worth noting that the onset crystallization of the *bcc* phase determined by DSC measurements increased from 421 °C with a Cr content at 10% to 456 °C at 30%, indicating that *bcc* phase formed at higher temperatures with an increasing Cr content. On the other hand, the crystallization temperature for boride phases was decreased with an increasing B content. This indicated that with a high B content, the limited solubility of B in the *bcc* phase saturated at a relatively low temperature, and thus promoted the boron segregation. In general a reduced B content is required as increasing the Cr content to avoid the overlapping of the onset crystallization temperatures for *bcc* and boride phases.

Similar crystallization process was observed in CoFeCrB films. Fig. 3 shows the *2θ* scans taken in the sample with a composition of $Co_{40}Fe_{20}Cr_{30}B_{10}$ annealed at a temperature ranging from 350 °C to 450 °C. The as-deposited film showed an amorphous feature. The *bcc* phase textured in (110) orientation emerged from the amorphous matrix at a temperature of 400 °C. The lattice constant was estimated ~ 2.86 Å according to the (110) peak at 44.94°, consistent with that of the ribbon samples. As the annealing temperature was raised up to 450 °C, a small peak corresponding to the boride-type phase $(Fe,Co)_3B$ was detected, as observed in ribbon materials. It was also noticed that the two onset crystallization temperatures for *bcc* and boride phases in ribbon samples with the same composition were 456°C and 573°C respectively, higher than those of the thin film samples.

**B. Magnetic properties**

Table II shows the experimental magnetizations $M_{EP}$ of as-deposited thin films and as-spun ribbon samples with different compositions, which are also plotted in Fig. 4 as a function of the Cr content in solid black square and solid red circle respectively. The magnetization was reduced with an increased Cr content, consistent with previous reports[8,9]. In as-deposited films the magnetization was ~ 651 emu/cc (0.83 $\mu_B$) with a Cr content of 10%, lower than the typical value obtained in $Co_{40}Fe_{40}B_{20}$. As the Cr content was increased to 30%, the magnetization was close to 0. The magnetization of Co, Fe, Ni based alloys could be estimated using the generalized Slater-Pauling curve[9,15,16]. In particular, Ref. 9 gave the calculated magnetic moments for each elements based on the generalized Slater-Pauling curve (2.6 $\mu_B$ for Fe, 1.6 $\mu_B$ for Co, -5.4 $\mu_B$ for Cr, and -2.4 $\mu_B$ for B). The trend of the experiment magnetizations in our study was well-described with these calculated magnetic moments. However, there is a large divergence between the calculated values and measured magnetic moments. The calculated moment was on average shifted down by ~0.54 $\mu_B$ for thin film samples, and ~0.64 $\mu_B$ for ribbon samples. For samples with 30% of Cr, the calculated magnetic moments are -0.8 $\mu_B$ (thin films) and -0.7 $\mu_B$ (ribbons) suggesting significant deviation in the value of the magnetic moment using the generalized Slater-Pauling curve.

The generalized Slater-Pauling curve shows fairly good agreement with experiment values of Fe/Co moments. Therefore, in our study, $Co_{40}Fe_{40}B_{20}$ is used as a standard sample to estimate the magnetic contribution of B atoms. Our ribbon sample $Co_{40}Fe_{40}B_{20}$ gave a magnetization of 1.49 $\mu_B$. Accordingly the average magnetic contribution from B is thus -0.94 $\mu_B$ instead of -2.4 $\mu_B$ calculated using the generalized Slater-Pauling curve. The discrepancy could be ascribed to the under-estimation of the magnetic contribution from the $s$ conduction band $2N_s^\uparrow$ for B atoms[17]. In most cases, $2N_s^\uparrow$ is assumed as a constant 0.6[9,15]. However, it has been pointed out in Ref. 16 that the s band contribution $2N_s^\uparrow$ for metalloid element B is as large as 1.85, which gave an effective B moment around -1.15 $\mu_B$, close to our estimation of -0.94 $\mu_B$. Additionally, the fittings for both thin film and ribbon samples could be further improved with a reduced Cr magnetic moment, which was similarly pointed out in Ref. 9 and was ascribed to the drastic assumption of $N_d^\uparrow = 0$ for Cr atoms. Besides, the divergence from the calculated magnetic moment of Cr atoms could be due to the fact that the generalized Slater-Pauling curve does not take the local atomic environment into account. The black dash line in Fig. 4 represents the calculated magnetizations $M_C$ of thin film samples with modified magnetic moments for B (-0.94 $\mu_B$) and Cr (-3.6 $\mu_B$). In addition, the experiment magnetizations for the ribbon samples were also well reproduced using the generalized Slater-Pauling curve with the same modified moments for B and Cr.

The effects of the annealing treatments on the magnetic properties of thin films samples were shown in Fig. 5. For each composition, the saturation magnetization ($M_s$) was increased by elevating the annealing temperature (Fig. 5(a)). This could be attributed to the fact that the smaller size of B atoms led to a more efficient diffusion compared to that of other atoms. The reduction of the B content resulted in an improved film magnetization, which was similarly observed in CoFeB[18].

Fig. 5(b) shows the coercive field ($H_c$) as functions of the annealing temperature in CoFeCrB films with different Cr contents. With a Cr content ~ 10%, the coercive field was slightly increased up to ~ 70 Oe as the annealing temperature was elevated to 425 ˚C. In samples with a 30% Cr content, the enhancement of the coercive field became significant from ~0 to ~519 Oe when the annealing temperature was increased to 425 ˚C. The main enhancement of the coercive field cannot be attributed to boron segregation at high annealing temperatures. In our study, boride phase was observed in $Co_{40}Fe_{40}B_{20}$ ribbon samples annealed above 430 ˚C, but the coercive field only increased up to ~40 Oe after annealing at 450 ˚C. Instead the mechanism of the coercive field could be similar to Co-Fe-Cr systems[19-22]. With the Cr content ranged from 10% to 40%, Kaneko alloy Co-Fe-Cr has a spinodal decomposition from a *α-bcc* solid solution into a two-phase structure, in which the Fe rich *bcc* clusters ($α_1$-phase) were magnetically isolated in Cr rich *bcc* matrix ($α_2$-phase). The high coercive field could be ascribed to the domain wall pinning effect or incoherent rotation among the $α_1$-phase clusters[19,23,24]. Besides, it is worth noting that the coercive field showed a significant dependence on the annealing temperature. In Ref. 22, the coercive field of alloy CoFeCrV remained below 80 Oe at 630 ˚C and was improved to 520 Oe at 650 ˚C. Bright field TEM studies indicated that spherical $α_1$ particles at 630 ˚C evolved into elongated ones at 650 ˚C, and the shape anisotropy contributed to the dramatic enhancement in $H_c$[20,22,24]. Further investigations using HRTEM is necessary to confirm our speculation on the mechanism for the increase of coercive fields for CoFeCrB.

The Gilbert damping parameter was characterized for $Co_{43}Fe_{29}Cr_{10}B_{18}$ films[16]. In as-deposited samples the damping parameter was ~ 0.0085, and was reduced to 0.006 after annealing at 450 ˚C (no boride phase was detected at this temperature). The damping parameter was comparable with the value obtained in $Co_{40}Fe_{20}B_{20}$ films[4,25]. Besides, it was much lower than previously reported values (0.02-0.028) for $Co_{40}Fe_{40-x}Cr_xB_{20}$ films with the Cr content *x* varied in the range of 0 to ~18%[9]. This could be possibly due to the improved compositional homogeneity and less defects in the films in our study.

## Conclusion:

We investigated the structural and magnetic properties of CoFeCrB with the Cr content varied in the range of 0-30%. An amorphous phase was observed in as-spun ribbons and as-deposited thin films. In ribbon samples, it was confirmed by both XRD scans and DSC measurements that a *bcc* phase was formed at a relatively low annealing temperature followed by the boron segregation at a higher annealing temperature. The onset temperature for the *bcc* phase was raised with an increased Cr content, while a higher B content led to a lower onset temperature for the boride phase (Fe,Co)$_3$B. The same crystallization process was observed in thin film samples. The magnetization of CoFeCrB was significantly reduced with an increased Cr content. The correlation between the magnetization and the Cr content can be well described using the generalized Slater-Pauling curve with modified Cr and B moments. Additionally, the annealing treatments also led to an increasing magnetization in thin films, likely due to the boron diffusion. A large Cr content also significantly enhanced the coercive field of CoFeCrB thin films after annealing at a high temperature, and we attributed it to the spinodal decomposition of Fe-rich $α_1$-phase clusters from the Cr-rich $α_2$-phase matrix. The Gilbert damping parameter was estimated to be around 0.006 in Co$_{43}$Fe$_{29}$Cr$_{10}$B$_{18}$ films, which is comparable with the benchmark value of Co$_{40}$Fe$_{20}$B$_{20}$ films.

## Acknowledgement:

The authors gratefully acknowledge the financial support provided by Grandis Inc. through DARPA (Award number: HR0011-09-C-0023) and the NSF (Award number: ECCS-0948138).

**Figures & Figure Captions:**

Fig.1 (a) *2θ* scans and (b) Bright-field TEM image of as-spun $Co_{40}Fe_{32}Cr_{10}B_{18}$ ribbons. The inset of (b) shows the selected area electron diffraction pattern.

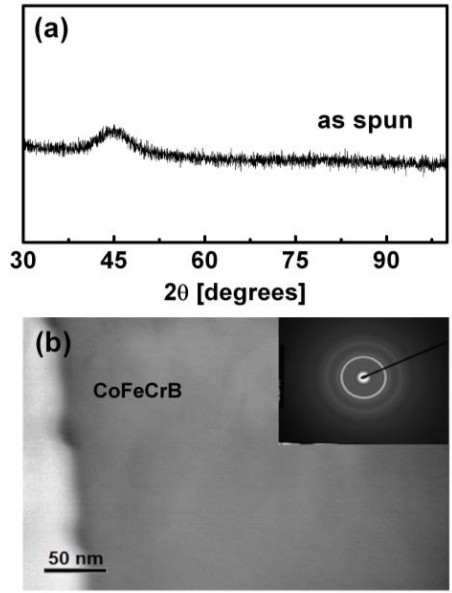

Fig.2 (Color online) (a) *2θ* scans of $Co_{40}Fe_{32}Cr_{10}B_{18}$ ribbons annealed at 417 ˚C and 508 ˚C respectively. (b) DSC curves of $Co_{40}Fe_{32}Cr_{10}B_{18}$ ribbons with the measurement temperature ramped from 50 ˚C to 600 ˚C.

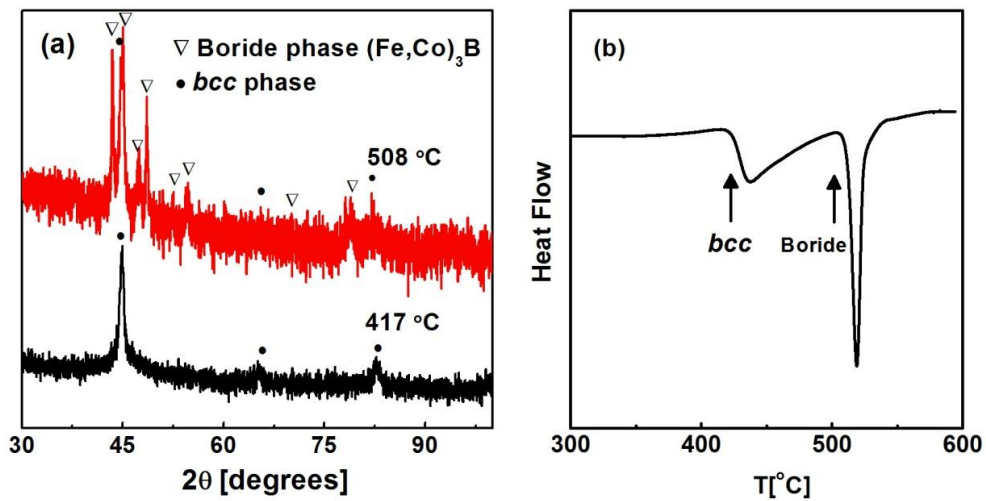

Fig.3 *2θ* scans of $Co_{40}Fe_{20}Cr_{30}B_{10}$ films annealed at a temperature ranging from 350 ˚C to 450 ˚C. Boride phase was observed in the sample annealed at 450 ˚C.

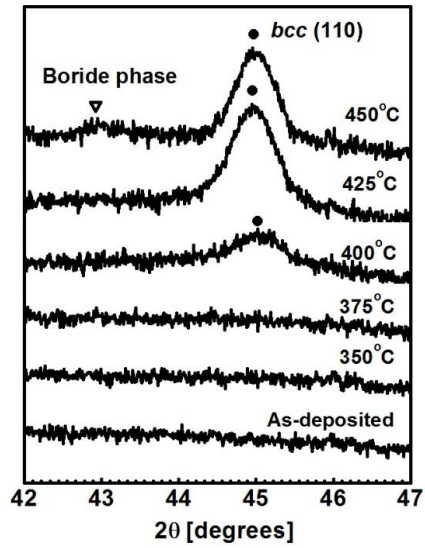

Fig.4 (Color online) Magnetizations of as-deposited films and as-spun ribbons varied with the Cr content. The black squares (red circles) stand for experiment values $M_{EP}$ obtained from as-deposited thin films (as-spun ribbons), the black (red) dash line represents the calculated values $M_C$ for the thin film samples (ribbon samples) using the generalized Slater-Pauling curve with modified magnetic moments for B (-0.94 $\mu_B$) and Cr (-3.6 $\mu_B$).

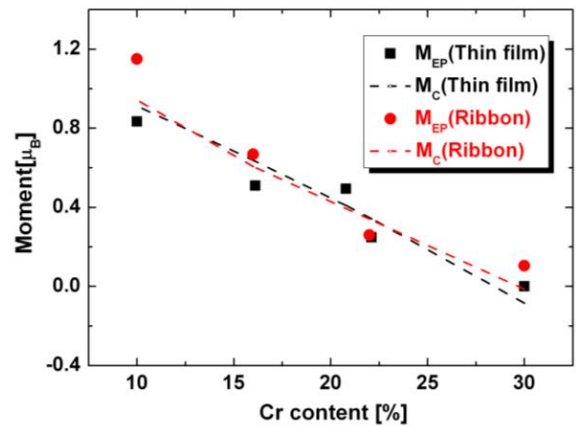

Fig.5 Magnetization $M_s$ (a) and Coercive field $H_c$ (b) of CoFeCrB thin films with different Cr contents. For each composition, characterizations were taken for as-deposited samples (squares) and samples annealed at 350 ˚C (circles), 375 ˚C (up triangles), 400 ˚C (down triangles), 425 ˚C (stars).

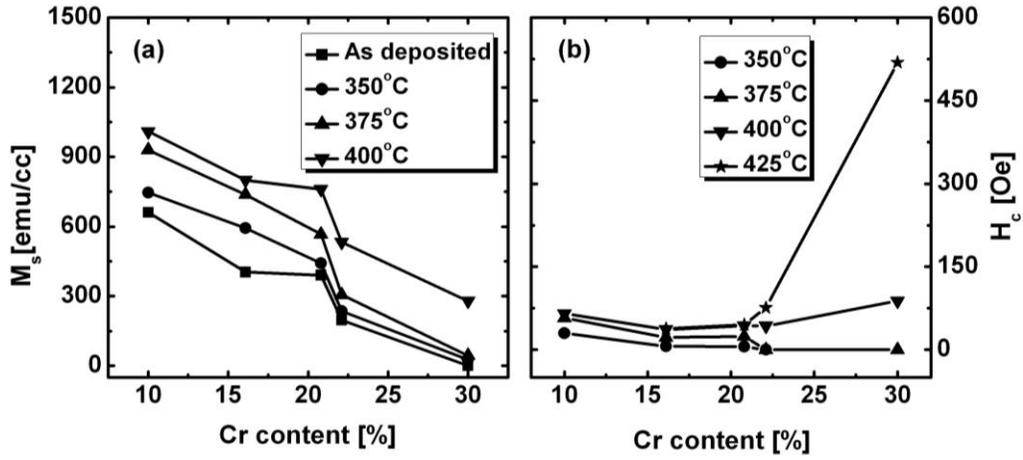

**Tables & Table Captions:**

Table I Ribbon sample compositions and the corresponding onset temperatures ($T_x$) of two crystallization phases determined by DSC measurements.

| Composition | Cr content (%) | B content (%) | Crystallization temperature | |
|---|---|---|---|---|
| | | | BCC $T_{x1}$ (°C) | Boride $T_{x2}$ (°C) |
| $Co_{40}Fe_{32}Cr_{10}B_{18}$ | 10 | 18 | 421 | 518 |
| $Co_{40}Fe_{27}Cr_{16}B_{17}$ | 16 | 17 | 426 | 541 |
| $Co_{40}Fe_{24}Cr_{22}B_{14}$ | 22 | 14 | 437 | 560 |
| $Co_{40}Fe_{20}Cr_{30}B_{10}$ | 30 | 10 | 456 | 573 |

Table II Compositions and corresponding magnetizations of as-deposited thin films and as-spun ribbons.

| Composition (Thin film) | Magnetization ($\mu_B$) | Composition (Ribbon) | Magnetization ($\mu_B$) |
|---|---|---|---|
| $Co_{43}Fe_{29}Cr_{10}B_{18}$ | 0.83 | $Co_{40}Fe_{32}Cr_{10}B_{18}$ | 1.15 |
| $Co_{44.8}Fe_{24.3}Cr_{16.1}B_{14.7}$ | 0.51 | $Co_{40}Fe_{27}Cr_{16}B_{17}$ | 0.67 |
| $Co_{44.6}Fe_{21.7}Cr_{20.8}B_{12.9}$ | 0.49 | $Co_{40}Fe_{24}Cr_{22}B_{14}$ | 0.26 |
| $Co_{44.6}Fe_{20.8}Cr_{22.1}B_{12.4}$ | 0.25 | $Co_{40}Fe_{20}Cr_{30}B_{10}$ | 0.10 |
| $Co_{40}Fe_{18}Cr_{30}B_{12}$ | ~0 | | |